\documentclass[sigchi-a]{acmart}

\usepackage{hyperref}
\hypersetup{
    pdftitle={The Potential of the Confluence of Theoretical and Algorithmic Modeling in Music Recommendation},
    pdflang={en-US},
    pdfauthor={Christine Bauer},
    pdfkeywords={computational modeling,music recommendation,preference prediction,human-computer interaction}
}

\def\BibTeX{{\rm B\kern-.05em{\sc i\kern-.025em b}\kern-.08emT\kern-.1667em\lower.7ex\hbox{E}\kern-.125emX}}
    
\copyrightyear{2019} 
\acmYear{2019} 
\setcopyright{rightsretained} 
\acmConference[CHI'19 Extended Abstracts]{CHI Conference on Human Factors in Computing Systems Extended Abstracts}{May 4--9, 2019}{Glasgow, Scotland UK}
\acmBooktitle{CHI Conference on Human Factors in Computing Systems Extended Abstracts (CHI'19 Extended Abstracts), May 4--9, 2019, Glasgow, Scotland UK}

\newif\ifworkinprogress
\workinprogresstrue

\ifworkinprogress
	\newcommand{\chb}[1]{\textcolor{magenta}{\textbf{[Christine] #1}}}
\else
  \newcommand{\chb}[1]{}
\fi

\begin{document}
\title{The Potential of the Confluence of Theoretical and Algorithmic Modeling in Music Recommendation}

\author{Christine Bauer}
\orcid{0000-0001-5724-1137}
\affiliation{%
  \institution{Johannes Kepler University Linz}
  \city{Linz} 
  \streetaddress{Altenberger Str. 69}
  \country{Austria} 
  \postcode{4040}
}
\email{christine.bauer@jku.at}

%

\begin{abstract}
The task of a music recommender system is to predict what music item a particular user would like to listen to next. 
%
This position paper discusses the main challenges of the music preference prediction task: the lack of information on the many contextual factors influencing a user's music preferences in existing open datasets, the lack of clarity of what the right choice of music is and whether a right choice exists at all; the multitude of criteria (beyond accuracy) that have to be met for a ``good'' music item recommendation; and the need for explanations on relationships to identify (and potentially counteract) unwanted biases in recommendation approaches.


The paper substantiates the position that the confluence of theoretical modeling (which seeks to explain behaviors) and algorithmic modeling (which seeks to predict behaviors) seems to be an effective avenue to take in computational modeling for music recommender systems.
\end{abstract}

%
%

%


%

%
\maketitle

\section{Introduction}

Before the era of the Internet, access to music content (e.g., music recordings) 
was restricted to local availability of their physical representations (e.g., vinyl). 
Thereby, the selection and aggregation of content had traditionally been exposed to human control~\cite{oechslein2014}. For instance, a small group of 
Artist\&Repertoire managers working for the major music labels scouted new artists and developed them commercially.

Nowadays,---owing to the development of the Social Web that allows for easy distribution of user-generated content---the intermediary level of experts (e.g., 
the Artist\&Repertoire managers at music labels) 
that traditionally ``prefiltered'' content before it reached potential consumers is bypassed. This results in the situation that users currently face: music content 
is abundantly available online and the amount of overall available content increases tremendously on a daily basis. 

\begin{sidebar}
    
    Examples of differences associated with music items, users, and their consumption behaviour include the following\cite{schedl2015_rshb}: 
    \begin{itemize}
        \item very low consumption time in the dimension of minutes, whereas a book or a travel are consumed during days or weeks; 
        \item consumption in sequences (e.g., playlists); 
        \item music often consumed passively (e.g., while jogging, travelling, working);  
        \item consumption is highly driven by situational context;
        \item users are  likely to appreciate the re-recommendation of the same item while a user is less likely to read the same news article over and over again; and
        \item music evokes strong emotions. 
    \end{itemize}
    
 \caption{The specialties of the music domain}
 \label{sidebar:specialties_music}
\end{sidebar}

However, the opportunity to access a large amount of content frequently leads to information overload~\cite{bawden2009} or choice overload~\cite{haubl2000consumer}, because people do not find the content that they are interested in or do not know what to choose.
Assisting users in searching, sorting, and filtering the massive amount of online content~\cite{montaner2003}, recommender systems have become important tools in people's everyday life and do not only facilitate the interaction with music content~\cite{celma2010_music_recommendation_and_discovery}, but also support versatile activities such as shopping~\cite{oestreicher2012}, consuming news~\cite{oechslein2014}, 
or finding persons for any kind of social matching~\cite{Mayer:2015:tois}.

Recommender systems are computer systems that 
provide suggestions for items 
that are 
deemed interesting to a particular target user, 
assisting that particular user in various decision-making processes (e.g., relating to what music to listen to)~\cite{ricci2015_rshb}. 
The general term used to denote to what the system recommends to users is ``item''~\cite{ricci2015_rshb}; 
in case of music recommender systems (MRS) it is the music item (e.g., musical work, artist, genre).

There are universally valid principles for designing recommender systems, such as 
that a recommender system typically consist of three key components (i.e., user, item, and matching mechanism)~\cite{bauer2017_bled_nonsuperstar_rs}. 
Still, a recommender system needs to be put into context because there are product- and sector-specific characteristics that a recommender system needs to consider (be customized to) to provide useful and effective recommendations for the specific type of item ~\cite{ricci2015_rshb,schedl2014_foundations}.
%
Sidebar~\ref{sidebar:specialties_music} presents the specialties of the music domain compared to other domains deploying recommender systems.

\section{Rationale}\label{rationale}

An ideal MRS proposes ``the right music, to the right user, at the right moment''~\cite{laplante2014}.
However, this is a complex task because various factors influence a user's music preferences in a given situation~\cite{bauer_schedl:hicss:2018}. 
Many studies have investigated the relationships between music preferences and various person-related characteristics (e.g., demographics~\cite{Cheng:2017:EUI:3077136.3080772}, personality traits~\cite{schafer2017can}, social influences~\cite{BonnevilleRoussy2018_socialinfluences}. 
Besides person-related characteristics, also situation-related factors (e.g., 
temporal aspects~\cite{dias2013improving}, or weather~\cite{braunhofer2011recommending}) influence a user's music preferences.
The task of an MRS is to predict what a particular user would like to listen to next. 
Basically, there are two computational modeling approaches to build upon for this music preference prediction task:

\begin{itemize}
	\item \emph{Theoretical modeling} seeks to explain users' listening behavior. For advancing MRS, the first step would be to observe a user's listening behavior and perform analyses to explain where a user's listening behavior results from (e.g., from person-related characteristics or situational factors, and from which of these in particular). Then, building on these findings (e.g., knowing that Finnish listeners are more likely to prefer heavy metal than Italian listeners~\cite{ijmir:schedl:2017}), future user models may be created for predictions.
	\item \emph{Algorithmic modeling} seeks to predict users' listening behavior. 
	Algorithmic modeling may rely on approaches that are capable of identifying listening patterns within a user's listening history or across users without necessarily delivering descriptions that help \emph{explaining} the relationships 
	of the identified patterns. For instance, approaches such as deep neural networks 
	frequently leave us with ``black boxes''~\cite{Koh:2017:UBP:3305381.3305576} 
	because the resulting models are complex and frequently they do not produce an intelligible description of the results produced in each case. Still, the resulting models may be apt to deliver remarkably accurate predictions. In other words, algorithmic modeling may recommend music to the user what he or she will indeed like in the very moment without understanding whether it was indeed the ``right'' choice---and if---why it was ``right''.
\end{itemize}

\section{Challenges}\label{challenges}
One challenge for music preference prediction is that it is (almost) impossible to say what is \emph{the} right choice for a particular user in the particular moment; it is typically a set of items that is \emph{right} or \emph{okay}.

Another challenge of algorithmic modeling is that---currently---we can only model based on data that we have available. For MRS, several open datasets exist, such as the Million Song Dataset~\cite{Bertin-Mahieux2011_millionsongdataset}, the LFM-1b dataset~\cite{schedl:icmr:2016}, or the recently released Music Streaming Sessions Dataset~\cite{brost2019music_spotifyskipdataset}.
However, there are many factors influencing a user's music preferences for which we do not have (sufficient) data available (yet) to exploit for algorithmic modeling. Theoretical modeling---thus, the ``explaining approach''---may help here to advance MRS. It is also a viable basis to provide an informed route what kind of data should be collected so that algorithmic modeling may come into play here to use its powerful mechanisms to exploit the additional data to make even better predictions.

A further challenge relates to evaluation of MRS: What does it mean if an MRS recommends a music item to a user and the user indeed listens to the item? Potentially, it is the user's most favorite song and so the user enjoyed listening to it. Maybe, though, the user listens to the item because the algorithm provided it as the next one to listen to in the playlist, but the user was distracted at the very moment because of receiving a phone call (or was not present in the room for some minutes). In such cases, the recommendation was maybe not a ``bad'' one because the user did not hear it anyways, but was it a good prediction then?

With respect to biases as inherent in recommendation systems (e.g., the popularity bias phenomenon~\cite{Celma:2008_popularity_bias} suggesting that over time the most popular music items tend to get more and more attention, while music items in the long tail get less and less attention~\cite{levy2010music}), the ability to understand and explain models seems to be a crucial prerequisite to uncover such bias and develop and take effective measures to counteract unwanted bias.

\section{Previous and Ongoing Research, and Interests}\label{sec:interests}

A major part of my previous and ongoing research is aimed at integrating contextual information into (user) modeling. Basically, my work on context modeling takes a conceptual viewpoint (e.g.,~\cite{bauer_novotny2017_consolidatedview,bauer2014_ccfis}). It points towards the various potentially relevant contextual factors that we tend to ``forget'' in modeling (for various reasons such as, for instance, the non-availability of useful datasets including such contextual information).

With the main objective at improving MRS, some part of my research on MRS is geared towards identifying relationships between various aspects (such as age~\cite{schedl_bauer2017_kidrec}, user connections~\cite{bauer_schedl_hicss2019_userconnections,bauer_schedl2018_ismir_userconnections}, user country~\cite{bauer_schedl:hicss:2018}, real-world events~\cite{schedl_wiechert_bauer2018_realworldevents}, mainstreaminess~\cite{schedl_bauer:momm:2017}) and music preferences or listening behavior.
Findings are then used to improve MRS performance (for instance in~\cite{schedl_bauer2017_kidrec,bauer_schedl:hicss:2018,schedl_bauer:momm:2017}).

To a considerable extent, ideas on the (contextual) components that could improve MRS are based on literature from various disciplines such as cognitive science (e.g.,~\cite{stevens2012_musicperception}), social psychology (e.g.,~\cite{bonneville2013_musicpreferences_adolescence}), and computer science~\cite{laplante2014}. In addition, ideas emanate from my own experience of many years in the music domain---which is a significant knowledge source that is not available to every researcher. 

\section{Prospects}\label{sec:prospects}

Overall, recommender systems research has predominantly focused on improving the prediction accuracy of algorithms based on existing datasets (reflecting users' historic item ratings or consumption behavior)~\cite{Beel:2013:online_offline}. 
However, to date, comprehensive contextual information about users and the specific situational settings in which those consume the items is rarely available in existing datasets~\cite{adomavicius2015_cars_handbook}---and is especially true for music-related datasets.

The confluence of theoretical and algorithmic modeling seems to be an effective avenue to take in computational modeling for MRS.


\begin{margintable}
\flushleft
\textbf{ACKNOWLEDGMENTS}\\
This research is supported by the \grantsponsor{fwf}{Austrian Science Fund 
 }: \grantnum[
 ]{fwf}{V579}.
\end{margintable}

\bibliographystyle{ACM-Reference-Format}
\bibliography{chi2019_hcicompmodeling}


\begin{thebibliography}{36}


\ifx \showCODEN    \undefined \def \showCODEN     #1{\unskip}     \fi
\ifx \showDOI      \undefined \def \showDOI       #1{#1}\fi
\ifx \showISBNx    \undefined \def \showISBNx     #1{\unskip}     \fi
\ifx \showISBNxiii \undefined \def \showISBNxiii  #1{\unskip}     \fi
\ifx \showISSN     \undefined \def \showISSN      #1{\unskip}     \fi
\ifx \showLCCN     \undefined \def \showLCCN      #1{\unskip}     \fi
\ifx \shownote     \undefined \def \shownote      #1{#1}          \fi
\ifx \showarticletitle \undefined \def \showarticletitle #1{#1}   \fi
\ifx \showURL      \undefined \def \showURL       {\relax}        \fi
\providecommand\bibfield[2]{#2}
\providecommand\bibinfo[2]{#2}
\providecommand\natexlab[1]{#1}
\providecommand\showeprint[2][]{arXiv:#2}

\bibitem[\protect\citeauthoryear{Adomavicius and Tuzhilin}{Adomavicius and
  Tuzhilin}{2015}]%
        {adomavicius2015_cars_handbook}
\bibfield{author}{\bibinfo{person}{Gediminas Adomavicius} {and}
  \bibinfo{person}{Alexander Tuzhilin}.} \bibinfo{year}{2015}\natexlab{}.
\newblock \showarticletitle{Context-aware recommender systems}.
\newblock In \bibinfo{booktitle}{\emph{Recommender Systems Handbook}
  (\bibinfo{edition}{2nd} ed.)}, \bibfield{editor}{\bibinfo{person}{Francesco
  Ricci}, \bibinfo{person}{Lior Rokach}, {and} \bibinfo{person}{Bracha
  Shapira}} (Eds.). \bibinfo{publisher}{Springer}, \bibinfo{address}{New York,
  NY}, \bibinfo{pages}{191--226}.
\newblock
\urldef\tempurl%
\url{https://doi.org/10.1007/978-1-4899-7637-6_6}
\showDOI{\tempurl}


\bibitem[\protect\citeauthoryear{Bauer}{Bauer}{2014}]%
        {bauer2014_ccfis}
\bibfield{author}{\bibinfo{person}{Christine Bauer}.}
  \bibinfo{year}{2014}\natexlab{}.
\newblock \showarticletitle{A framework for conceptualizing context for
  intelligent systems (CCFIS)}.
\newblock \bibinfo{journal}{\emph{Journal of Ambient Intelligence and Smart
  Environments}} \bibinfo{volume}{6}, \bibinfo{number}{4}
  (\bibinfo{year}{2014}), \bibinfo{pages}{403--417}.
\newblock
\showISSN{1876-1364}
\urldef\tempurl%
\url{https://doi.org/10.3233/AIS-140269}
\showDOI{\tempurl}


\bibitem[\protect\citeauthoryear{Bauer, Kholodylo, and Strauss}{Bauer
  et~al\mbox{.}}{2017}]%
        {bauer2017_bled_nonsuperstar_rs}
\bibfield{author}{\bibinfo{person}{Christine Bauer}, \bibinfo{person}{Marta
  Kholodylo}, {and} \bibinfo{person}{Christine Strauss}.}
  \bibinfo{year}{2017}\natexlab{}.
\newblock \showarticletitle{Music Recommender Systems: Challenges and
  Opportunities for Non-Superstar Artists}. In \bibinfo{booktitle}{\emph{30th
  Bled eConference}}. \bibinfo{pages}{21--32}.
\newblock
\showISBNx{978-961-286-043-1}
\urldef\tempurl%
\url{https://doi.org/10.18690/978-961-286-043-1.3}
\showDOI{\tempurl}


\bibitem[\protect\citeauthoryear{Bauer and Novotny}{Bauer and Novotny}{2017}]%
        {bauer_novotny2017_consolidatedview}
\bibfield{author}{\bibinfo{person}{Christine Bauer} {and}
  \bibinfo{person}{Alexander Novotny}.} \bibinfo{year}{2017}\natexlab{}.
\newblock \showarticletitle{A consolidated view of context for intelligent
  systems}.
\newblock \bibinfo{journal}{\emph{Journal of Ambient Intelligence and Smart
  Environments}} \bibinfo{volume}{9}, \bibinfo{number}{4}
  (\bibinfo{year}{2017}), \bibinfo{pages}{377--393}.
\newblock
\showISSN{18761372, 18761364}
\urldef\tempurl%
\url{https://doi.org/10.3233/ais-170445}
\showDOI{\tempurl}


\bibitem[\protect\citeauthoryear{Bauer and Schedl}{Bauer and Schedl}{2018a}]%
        {bauer_schedl2018_ismir_userconnections}
\bibfield{author}{\bibinfo{person}{Christine Bauer} {and}
  \bibinfo{person}{Markus Schedl}.} \bibinfo{year}{2018}\natexlab{a}.
\newblock \showarticletitle{Investigating cross-country relationship between
  users' social ties and music mainstreaminess}. In
  \bibinfo{booktitle}{\emph{19th International Society for Music Information
  Retrieval Conference}} \emph{(\bibinfo{series}{ISMIR`18})}.
  \bibinfo{publisher}{ISMIR}, \bibinfo{pages}{678--686}.
\newblock


\bibitem[\protect\citeauthoryear{Bauer and Schedl}{Bauer and Schedl}{2018b}]%
        {bauer_schedl:hicss:2018}
\bibfield{author}{\bibinfo{person}{Christine Bauer} {and}
  \bibinfo{person}{Markus Schedl}.} \bibinfo{year}{2018}\natexlab{b}.
\newblock \showarticletitle{On the Importance of Considering Country-specific
  Aspects on the Online-Market: An Example of Music Recommendation Considering
  Country-Specific Mainstream}. In \bibinfo{booktitle}{\emph{51st Hawaii
  International Conference on System Sciences}}
  \emph{(\bibinfo{series}{HICSS`18})}. \bibinfo{pages}{3647--3656}.
\newblock
\showISBNx{978-0-9981331-1-9}
\urldef\tempurl%
\url{http://hdl.handle.net/10125/50349}
\showURL{%
\tempurl}


\bibitem[\protect\citeauthoryear{Bauer and Schedl}{Bauer and Schedl}{2019}]%
        {bauer_schedl_hicss2019_userconnections}
\bibfield{author}{\bibinfo{person}{Christine Bauer} {and}
  \bibinfo{person}{Markus Schedl}.} \bibinfo{year}{2019}\natexlab{}.
\newblock \showarticletitle{A cross-country investigation of user connection
  patterns in online social networks}. In \bibinfo{booktitle}{\emph{52nd Hawaii
  International Conference on System Sciences}}
  \emph{(\bibinfo{series}{HICSS`19})}. \bibinfo{pages}{2166--2175}.
\newblock
\showISBNx{978-0-9981331-2-6}
\urldef\tempurl%
\url{http://hdl.handle.net/10125/59655}
\showURL{%
\tempurl}


\bibitem[\protect\citeauthoryear{Bawden and Robinson}{Bawden and
  Robinson}{2009}]%
        {bawden2009}
\bibfield{author}{\bibinfo{person}{David Bawden} {and} \bibinfo{person}{Lyn
  Robinson}.} \bibinfo{year}{2009}\natexlab{}.
\newblock \showarticletitle{The dark side of information: overload, anxiety and
  other paradoxes and pathologies}.
\newblock \bibinfo{journal}{\emph{Journal of Information Science}}
  \bibinfo{volume}{35}, \bibinfo{number}{2} (\bibinfo{year}{2009}),
  \bibinfo{pages}{180--191}.
\newblock
\showISSN{0165-5515, 1741-6485}
\urldef\tempurl%
\url{https://doi.org/10.1177/0165551508095781}
\showDOI{\tempurl}


\bibitem[\protect\citeauthoryear{Beel, Genzmehr, Langer, N\"{u}rnberger, and
  Gipp}{Beel et~al\mbox{.}}{2013}]%
        {Beel:2013:online_offline}
\bibfield{author}{\bibinfo{person}{Joeran Beel}, \bibinfo{person}{Marcel
  Genzmehr}, \bibinfo{person}{Stefan Langer}, \bibinfo{person}{Andreas
  N\"{u}rnberger}, {and} \bibinfo{person}{Bela Gipp}.}
  \bibinfo{year}{2013}\natexlab{}.
\newblock \showarticletitle{A Comparative Analysis of Offline and Online
  Evaluations and Discussion of Research Paper Recommender System Evaluation}.
  In \bibinfo{booktitle}{\emph{International Workshop on Reproducibility and
  Replication in Recommender Systems Evaluation}}
  \emph{(\bibinfo{series}{RepSys`13})}. \bibinfo{publisher}{ACM},
  \bibinfo{pages}{7--14}.
\newblock
\showISBNx{978-1-4503-2465-6}
\urldef\tempurl%
\url{https://doi.org/10.1145/2532508.2532511}
\showDOI{\tempurl}


\bibitem[\protect\citeauthoryear{Bertin-Mahieux, Ellis, Whitman, and
  Lamere}{Bertin-Mahieux et~al\mbox{.}}{2011}]%
        {Bertin-Mahieux2011_millionsongdataset}
\bibfield{author}{\bibinfo{person}{Thierry Bertin-Mahieux},
  \bibinfo{person}{Daniel~P.W. Ellis}, \bibinfo{person}{Brian Whitman}, {and}
  \bibinfo{person}{Paul Lamere}.} \bibinfo{year}{2011}\natexlab{}.
\newblock \showarticletitle{The Million Song Dataset}. In
  \bibinfo{booktitle}{\emph{12th International Conference on Music Information
  Retrieval}} \emph{(\bibinfo{series}{ISMIR`11})}. \bibinfo{publisher}{ISMIR}.
\newblock


\bibitem[\protect\citeauthoryear{Bonneville-Roussy, Rentfrow, Xu, and
  Potter}{Bonneville-Roussy et~al\mbox{.}}{2013}]%
        {bonneville2013_musicpreferences_adolescence}
\bibfield{author}{\bibinfo{person}{Arielle Bonneville-Roussy},
  \bibinfo{person}{Peter~J. Rentfrow}, \bibinfo{person}{Man~K. Xu}, {and}
  \bibinfo{person}{Jeff Potter}.} \bibinfo{year}{2013}\natexlab{}.
\newblock \showarticletitle{Music through the ages: Trends in musical
  engagement and preferences from adolescence through middle adulthood}.
\newblock \bibinfo{journal}{\emph{Journal of Personality and Social
  Psychology}} \bibinfo{volume}{105}, \bibinfo{number}{4}
  (\bibinfo{year}{2013}), \bibinfo{pages}{703--717}.
\newblock
\urldef\tempurl%
\url{https://doi.org/10.1037/a0033770}
\showDOI{\tempurl}


\bibitem[\protect\citeauthoryear{Bonneville-Roussy and Rust}{Bonneville-Roussy
  and Rust}{2018}]%
        {BonnevilleRoussy2018_socialinfluences}
\bibfield{author}{\bibinfo{person}{Arielle Bonneville-Roussy} {and}
  \bibinfo{person}{John Rust}.} \bibinfo{year}{2018}\natexlab{}.
\newblock \showarticletitle{Age trends in musical preferences in adulthood: 2.
  Sources of social influences as determinants of preferences}.
\newblock \bibinfo{journal}{\emph{Musicae Scientiae}} \bibinfo{volume}{22},
  \bibinfo{number}{2} (\bibinfo{year}{2018}), \bibinfo{pages}{175--195}.
\newblock
\urldef\tempurl%
\url{https://doi.org/10.1177/1029864917704016}
\showDOI{\tempurl}


\bibitem[\protect\citeauthoryear{Braunhofer, Kaminskas, and Ricci}{Braunhofer
  et~al\mbox{.}}{2011}]%
        {braunhofer2011recommending}
\bibfield{author}{\bibinfo{person}{Matthias Braunhofer},
  \bibinfo{person}{Marius Kaminskas}, {and} \bibinfo{person}{Francesco Ricci}.}
  \bibinfo{year}{2011}\natexlab{}.
\newblock \showarticletitle{Recommending music for places of interest in a
  mobile travel guide}. In \bibinfo{booktitle}{\emph{5th ACM Conference on
  Recommender Systems}} \emph{(\bibinfo{series}{RecSys`11})}. ACM,
  \bibinfo{pages}{253--256}.
\newblock


\bibitem[\protect\citeauthoryear{Brost, Mehrotra, and Jehan}{Brost
  et~al\mbox{.}}{2019}]%
        {brost2019music_spotifyskipdataset}
\bibfield{author}{\bibinfo{person}{Brian Brost}, \bibinfo{person}{Rishabh
  Mehrotra}, {and} \bibinfo{person}{Tristan Jehan}.}
  \bibinfo{year}{2019}\natexlab{}.
\newblock \showarticletitle{The Music Streaming Sessions Dataset}. In
  \bibinfo{booktitle}{\emph{The Web Conference 2019}}. ACM.
\newblock


\bibitem[\protect\citeauthoryear{Celma}{Celma}{2010}]%
        {celma2010_music_recommendation_and_discovery}
\bibfield{author}{\bibinfo{person}{\`{O}scar Celma}.}
  \bibinfo{year}{2010}\natexlab{}.
\newblock \bibinfo{booktitle}{\emph{Music Recommendation and Discovery: The
  Long Tail, Long Fail, and Long Play in the Digital Music Space}}.
\newblock \bibinfo{publisher}{Springer}, \bibinfo{address}{Berlin, Heidelberg,
  Germany}.
\newblock


\bibitem[\protect\citeauthoryear{Celma and Cano}{Celma and Cano}{2008}]%
        {Celma:2008_popularity_bias}
\bibfield{author}{\bibinfo{person}{\`{O}scar Celma} {and}
  \bibinfo{person}{Pedro Cano}.} \bibinfo{year}{2008}\natexlab{}.
\newblock \showarticletitle{From Hits to Niches?: Or How Popular Artists Can
  Bias Music Recommendation and Discovery}. In \bibinfo{booktitle}{\emph{2Nd
  KDD Workshop on Large-Scale Recommender Systems and the Netflix Prize
  Competition}} \emph{(\bibinfo{series}{NETFLIX`08})}.
  \bibinfo{publisher}{ACM}, \bibinfo{address}{New York, NY}, Article
  \bibinfo{articleno}{5}, \bibinfo{numpages}{8}~pages.
\newblock
\showISBNx{978-1-60558-265-8}
\urldef\tempurl%
\url{https://doi.org/10.1145/1722149.1722154}
\showDOI{\tempurl}


\bibitem[\protect\citeauthoryear{Cheng, Shen, Nie, Chua, and Kankanhalli}{Cheng
  et~al\mbox{.}}{2017}]%
        {Cheng:2017:EUI:3077136.3080772}
\bibfield{author}{\bibinfo{person}{Zhiyong Cheng}, \bibinfo{person}{Jialie
  Shen}, \bibinfo{person}{Liqiang Nie}, \bibinfo{person}{Tat-Seng Chua}, {and}
  \bibinfo{person}{Mohan Kankanhalli}.} \bibinfo{year}{2017}\natexlab{}.
\newblock \showarticletitle{Exploring User-Specific Information in Music
  Retrieval}. In \bibinfo{booktitle}{\emph{40th International ACM SIGIR
  Conference on Research and Development in Information Retrieval}}
  \emph{(\bibinfo{series}{SIGIR`17})}. \bibinfo{publisher}{ACM},
  \bibinfo{address}{New York, NY}, \bibinfo{pages}{655--664}.
\newblock
\showISBNx{978-1-4503-5022-8}
\urldef\tempurl%
\url{https://doi.org/10.1145/3077136.3080772}
\showDOI{\tempurl}


\bibitem[\protect\citeauthoryear{Dias and Fonseca}{Dias and Fonseca}{2013}]%
        {dias2013improving}
\bibfield{author}{\bibinfo{person}{Ricardo Dias} {and}
  \bibinfo{person}{Manuel~J Fonseca}.} \bibinfo{year}{2013}\natexlab{}.
\newblock \showarticletitle{Improving music recommendation in session-based
  collaborative filtering by using temporal context}. In
  \bibinfo{booktitle}{\emph{IEEE 25th International Conference on Tools with
  Artificial Intelligence}} \emph{(\bibinfo{series}{ICTAI`13})}. IEEE,
  \bibinfo{pages}{783--788}.
\newblock
\showISSN{1082-3409}
\urldef\tempurl%
\url{https://doi.org/10.1109/ICTAI.2013.120}
\showDOI{\tempurl}


\bibitem[\protect\citeauthoryear{H{\"a}ubl and Trifts}{H{\"a}ubl and
  Trifts}{2000}]%
        {haubl2000consumer}
\bibfield{author}{\bibinfo{person}{Gerald H{\"a}ubl} {and}
  \bibinfo{person}{Valerie Trifts}.} \bibinfo{year}{2000}\natexlab{}.
\newblock \showarticletitle{Consumer decision making in online shopping
  environments: The effects of interactive decision aids}.
\newblock \bibinfo{journal}{\emph{Marketing Science}} \bibinfo{volume}{19},
  \bibinfo{number}{1} (\bibinfo{year}{2000}), \bibinfo{pages}{4--21}.
\newblock


\bibitem[\protect\citeauthoryear{Koh and Liang}{Koh and Liang}{2017}]%
        {Koh:2017:UBP:3305381.3305576}
\bibfield{author}{\bibinfo{person}{Pang~Wei Koh} {and} \bibinfo{person}{Percy
  Liang}.} \bibinfo{year}{2017}\natexlab{}.
\newblock \showarticletitle{Understanding Black-box Predictions via Influence
  Functions}. In \bibinfo{booktitle}{\emph{34th International Conference on
  Machine Learning}} \emph{(\bibinfo{series}{ICML`17})}.
  \bibinfo{pages}{1885--1894}.
\newblock


\bibitem[\protect\citeauthoryear{Laplante}{Laplante}{2014}]%
        {laplante2014}
\bibfield{author}{\bibinfo{person}{Audrey Laplante}.}
  \bibinfo{year}{2014}\natexlab{}.
\newblock \showarticletitle{Improving music recommender systems: what can we
  learn from research on music tags?}. In \bibinfo{booktitle}{\emph{15th
  International Society for Music Information Retrieval Conference}}
  \emph{(\bibinfo{series}{ISMIR`14})}. \bibinfo{publisher}{ISMIR},
  \bibinfo{pages}{451--456}.
\newblock


\bibitem[\protect\citeauthoryear{Levy and Bosteels}{Levy and Bosteels}{2010}]%
        {levy2010music}
\bibfield{author}{\bibinfo{person}{Mark Levy} {and} \bibinfo{person}{Klaas
  Bosteels}.} \bibinfo{year}{2010}\natexlab{}.
\newblock \showarticletitle{Music recommendation and the long tail}. In
  \bibinfo{booktitle}{\emph{1st Workshop On Music Recommendation And
  Discovery}} \emph{(\bibinfo{series}{WOMRAD`10})}.
\newblock


\bibitem[\protect\citeauthoryear{Mayer, Jones, and Hiltz}{Mayer
  et~al\mbox{.}}{2015}]%
        {Mayer:2015:tois}
\bibfield{author}{\bibinfo{person}{Julia~M. Mayer}, \bibinfo{person}{Quentin
  Jones}, {and} \bibinfo{person}{Starr~Roxanne Hiltz}.}
  \bibinfo{year}{2015}\natexlab{}.
\newblock \showarticletitle{Identifying Opportunities for Valuable Encounters:
  Toward Context-Aware Social Matching Systems}.
\newblock \bibinfo{journal}{\emph{ACM Transactions on Information Systems}}
  \bibinfo{volume}{34}, \bibinfo{number}{1}, Article \bibinfo{articleno}{1}
  (\bibinfo{date}{July} \bibinfo{year}{2015}), \bibinfo{numpages}{32}~pages.
\newblock
\showISSN{1046-8188}
\urldef\tempurl%
\url{https://doi.org/10.1145/2751557}
\showDOI{\tempurl}


\bibitem[\protect\citeauthoryear{Montaner, L\'{o}pez, and de~la Rosa}{Montaner
  et~al\mbox{.}}{2003}]%
        {montaner2003}
\bibfield{author}{\bibinfo{person}{Miquel Montaner}, \bibinfo{person}{Beatriz
  L\'{o}pez}, {and} \bibinfo{person}{Josep~Llu\'{i}s de~la Rosa}.}
  \bibinfo{year}{2003}\natexlab{}.
\newblock \showarticletitle{A taxonomy of recommender agents on the Internet}.
\newblock \bibinfo{journal}{\emph{Artificial Intelligence Review}}
  \bibinfo{volume}{19}, \bibinfo{number}{4} (\bibinfo{year}{2003}),
  \bibinfo{pages}{285--330}.
\newblock
\showISSN{0269-2821}
\urldef\tempurl%
\url{https://doi.org/10.1023/A:1022850703159}
\showDOI{\tempurl}


\bibitem[\protect\citeauthoryear{Oechslein and Hess}{Oechslein and
  Hess}{2014}]%
        {oechslein2014}
\bibfield{author}{\bibinfo{person}{Oliver Oechslein} {and}
  \bibinfo{person}{Thomas Hess}.} \bibinfo{year}{2014}\natexlab{}.
\newblock \showarticletitle{The Value of a Recommendation: The Role of Social
  Ties in Social Recommender Systems}. In \bibinfo{booktitle}{\emph{47th Hawaii
  International Conference on System Science}}
  \emph{(\bibinfo{series}{HICSS`14})}. \bibinfo{publisher}{IEEE},
  \bibinfo{pages}{1864--1873}.
\newblock
\urldef\tempurl%
\url{https://doi.org/10.1109/HICSS.2014.235}
\showDOI{\tempurl}


\bibitem[\protect\citeauthoryear{Oestreicher-Singer and
  Sundararajan}{Oestreicher-Singer and Sundararajan}{2012}]%
        {oestreicher2012}
\bibfield{author}{\bibinfo{person}{Gal Oestreicher-Singer} {and}
  \bibinfo{person}{Arun Sundararajan}.} \bibinfo{year}{2012}\natexlab{}.
\newblock \showarticletitle{Recommendation Networks and the Long Tail of
  Electronic Commerce}.
\newblock \bibinfo{journal}{\emph{MIS Quarterly}} \bibinfo{volume}{36},
  \bibinfo{number}{1} (\bibinfo{year}{2012}), \bibinfo{pages}{65--83}.
\newblock
\showISSN{0276-7783}


\bibitem[\protect\citeauthoryear{Ricci, Rokach, and Shapira}{Ricci
  et~al\mbox{.}}{2015}]%
        {ricci2015_rshb}
\bibfield{author}{\bibinfo{person}{Francesco Ricci}, \bibinfo{person}{Lior
  Rokach}, {and} \bibinfo{person}{Bracha Shapira}.}
  \bibinfo{year}{2015}\natexlab{}.
\newblock \bibinfo{booktitle}{\emph{Recommender Systems Handbook}
  (\bibinfo{edition}{2nd} ed.)}.
\newblock \bibinfo{publisher}{Springer}, \bibinfo{address}{New York, NY}.
\newblock
\urldef\tempurl%
\url{https://doi.org/10.1007/978-1-4899-7637-6}
\showDOI{\tempurl}


\bibitem[\protect\citeauthoryear{Sch\"{a}fer and Mehlhorn}{Sch\"{a}fer and
  Mehlhorn}{2017}]%
        {schafer2017can}
\bibfield{author}{\bibinfo{person}{Thomas Sch\"{a}fer} {and}
  \bibinfo{person}{Claudia Mehlhorn}.} \bibinfo{year}{2017}\natexlab{}.
\newblock \showarticletitle{Can personality traits predict musical style
  preferences? A meta-analysis}.
\newblock \bibinfo{journal}{\emph{Personality and Individual Differences}}
  \bibinfo{volume}{116} (\bibinfo{year}{2017}), \bibinfo{pages}{265--273}.
\newblock


\bibitem[\protect\citeauthoryear{Schedl}{Schedl}{2016}]%
        {schedl:icmr:2016}
\bibfield{author}{\bibinfo{person}{Markus Schedl}.}
  \bibinfo{year}{2016}\natexlab{}.
\newblock \showarticletitle{{The LFM-1b Dataset for Music Retrieval and
  Recommendation}}. In \bibinfo{booktitle}{\emph{ACM on International
  Conference on Multimedia Retrieval}} \emph{(\bibinfo{series}{ICMR`16})}.
  \bibinfo{publisher}{ACM}, \bibinfo{address}{New York, NY},
  \bibinfo{pages}{103--110}.
\newblock
\showISBNx{978-1-4503-4359-6}
\urldef\tempurl%
\url{https://doi.org/10.1145/2911996.2912004}
\showDOI{\tempurl}


\bibitem[\protect\citeauthoryear{Schedl}{Schedl}{2017}]%
        {ijmir:schedl:2017}
\bibfield{author}{\bibinfo{person}{Markus Schedl}.}
  \bibinfo{year}{2017}\natexlab{}.
\newblock \showarticletitle{{Investigating country-specific music preferences
  and music recommendation algorithms with the LFM-1b dataset}}.
\newblock \bibinfo{journal}{\emph{{International Journal of Multimedia
  Information Retrieval}}} \bibinfo{volume}{6}, \bibinfo{number}{1}
  (\bibinfo{year}{2017}), \bibinfo{pages}{71--84}.
\newblock
\showISSN{2192-662X}
\urldef\tempurl%
\url{https://doi.org/10.1007/s13735-017-0118-y}
\showDOI{\tempurl}


\bibitem[\protect\citeauthoryear{Schedl and Bauer}{Schedl and Bauer}{2017a}]%
        {schedl_bauer:momm:2017}
\bibfield{author}{\bibinfo{person}{Markus Schedl} {and}
  \bibinfo{person}{Christine Bauer}.} \bibinfo{year}{2017}\natexlab{a}.
\newblock \showarticletitle{{Introducing Global and Regional Mainstreaminess
  for Improving Personalized Music Recommendation}}. In
  \bibinfo{booktitle}{\emph{15th International Conference on Advances in Mobile
  Computing \& Multimedia}} \emph{(\bibinfo{series}{MoMM`17})}.
  \bibinfo{publisher}{ACM}, \bibinfo{address}{New York, NY},
  \bibinfo{pages}{74--81}.
\newblock
\showISBNx{978-1-4503-5300-7}
\urldef\tempurl%
\url{https://doi.org/10.1145/3151848.3151849}
\showDOI{\tempurl}


\bibitem[\protect\citeauthoryear{Schedl and Bauer}{Schedl and Bauer}{2017b}]%
        {schedl_bauer2017_kidrec}
\bibfield{author}{\bibinfo{person}{Markus Schedl} {and}
  \bibinfo{person}{Christine Bauer}.} \bibinfo{year}{2017}\natexlab{b}.
\newblock \showarticletitle{Online Music Listening Culture of Kids and
  Adolescents: Listening Analysis and Music Recommendation Tailored to the
  Young}. In \bibinfo{booktitle}{\emph{1st International Workshop on Children
  and Recommender Systems}} \emph{(\bibinfo{series}{KidRec`17})}.
  \bibinfo{publisher}{ACM}, \bibinfo{address}{New York, NY}.
\newblock


\bibitem[\protect\citeauthoryear{Schedl, G\'{o}mez, and Urbano}{Schedl
  et~al\mbox{.}}{2014}]%
        {schedl2014_foundations}
\bibfield{author}{\bibinfo{person}{Markus Schedl}, \bibinfo{person}{Emilia
  G\'{o}mez}, {and} \bibinfo{person}{Juli\'{a}n Urbano}.}
  \bibinfo{year}{2014}\natexlab{}.
\newblock \showarticletitle{Music Information Retrieval: Recent Developments
  and Applications}.
\newblock \bibinfo{journal}{\emph{Foundations and Trends in Information
  Retrieval}} \bibinfo{volume}{8}, \bibinfo{number}{2-3}
  (\bibinfo{year}{2014}), \bibinfo{pages}{127--261}.
\newblock


\bibitem[\protect\citeauthoryear{Schedl, Knees, McFee, Bogdanov, and
  Kaminskas}{Schedl et~al\mbox{.}}{2015}]%
        {schedl2015_rshb}
\bibfield{author}{\bibinfo{person}{Markus Schedl}, \bibinfo{person}{Peter
  Knees}, \bibinfo{person}{Brian McFee}, \bibinfo{person}{Dmitry Bogdanov},
  {and} \bibinfo{person}{Marius Kaminskas}.} \bibinfo{year}{2015}\natexlab{}.
\newblock \showarticletitle{Music Recommender Systems}.
\newblock In \bibinfo{booktitle}{\emph{Recommender Systems Handbook}
  (\bibinfo{edition}{2nd} ed.)}, \bibfield{editor}{\bibinfo{person}{Francesco
  Ricci}, \bibinfo{person}{Lior Rokach}, \bibinfo{person}{Bracha Shapira},
  {and} \bibinfo{person}{Paul~B. Kantor}} (Eds.).
  \bibinfo{publisher}{Springer}, \bibinfo{address}{New York, NY},
  \bibinfo{pages}{453--492}.
\newblock
\urldef\tempurl%
\url{https://doi.org/10.1007/978-1-4899-7637-6_13}
\showDOI{\tempurl}


\bibitem[\protect\citeauthoryear{Schedl, Wiechert, and Bauer}{Schedl
  et~al\mbox{.}}{[n. d.]}]%
        {schedl_wiechert_bauer2018_realworldevents}
\bibfield{author}{\bibinfo{person}{Markus Schedl}, \bibinfo{person}{Eelco
  Wiechert}, {and} \bibinfo{person}{Christine Bauer}.} \bibinfo{year}{[n.
  d.]}\natexlab{}.
\newblock \showarticletitle{The effects of real-world events on music listening
  behavior: an intervention time series analysis}. In
  \bibinfo{booktitle}{\emph{WWW '18 Companion: The 2018 Web Conference
  Companion}} \emph{(\bibinfo{series}{WWW`18})}. \bibinfo{publisher}{ACM},
  \bibinfo{pages}{75--76}.
\newblock
\showISBNx{978-1-4503-5640-4}
\urldef\tempurl%
\url{https://doi.org/10.1145/3184558.3186936}
\showDOI{\tempurl}


\bibitem[\protect\citeauthoryear{Stevens}{Stevens}{2012}]%
        {stevens2012_musicperception}
\bibfield{author}{\bibinfo{person}{Catherine~J. Stevens}.}
  \bibinfo{year}{2012}\natexlab{}.
\newblock \showarticletitle{Music perception and cognition: a review of recent
  cross-cultural research}.
\newblock \bibinfo{journal}{\emph{Topics in Cognitive Science}}
  \bibinfo{volume}{4}, \bibinfo{number}{4} (\bibinfo{year}{2012}),
  \bibinfo{pages}{653--667}.
\newblock
\urldef\tempurl%
\url{https://doi.org/10.1111/j.1756-8765.2012.01215.x}
\showDOI{\tempurl}


\end{thebibliography}

\end{document}